\documentclass[useAMS,usenatbib]{mn2e}

\usepackage{graphicx}
\usepackage{times}%
\usepackage{natbib}
\usepackage{graphics}
\usepackage{epsfig}
\usepackage{array}
\usepackage{stfloats}
\usepackage{fixltx2e}
\usepackage{amsmath}

\newcommand{\ie}{{i.e.}}
\newcommand{\eg}{{e.g.}}
\newcommand{\gsim}{\,\lower2truept\hbox{${>\atop\hbox{\raise4truept\hbox{$\sim$}}}$}\,}

\def\eg{{\rm e.g.$\,$}}
\def\ie{{\rm i.e.$\,$}}

\newcommand{\be}{\begin{equation}}
\newcommand{\ee}{\end{equation}}
\newcommand{\bea}{\begin{eqnarray}}
\newcommand{\eea}{\end{eqnarray}}

\def\ltsima{$\; \buildrel < \over \sim \;$}
\def\simlt{\lower.5ex\hbox{\ltsima}}
\def\gtsima{$\; \buildrel > \over \sim \;$}
\def\simgt{\lower.5ex\hbox{\gtsima}}

\title[Bouncing dark energy and high-z massive clusters]{Early massive clusters and the bouncing coupled dark energy}

\author[M. Baldi]{Marco Baldi
\\Excellence Cluster Universe, Boltzmannstr.~2, D-85748 Garching, Germany
\\University Observatory, Ludwig-Maximillians University Munich, Scheinerstr. 1, D-81679 Munich, Germany}

\begin{document}


\pagerange{\pageref{firstpage}--\pageref{lastpage}} \pubyear{2011}

\maketitle

\label{firstpage}

\begin{abstract}
\ \\
The abundance of the most massive objects in the Universe at different epochs is a very sensitive probe of the
cosmic background evolution and of the growth history of density perturbations, and could provide a powerful tool to 
distinguish between a cosmological constant and a dynamical dark energy field. In particular, the recent detection of
very massive clusters of galaxies at high redshifts has attracted significant interest as a possible indication of a failure of the
standard $\Lambda $CDM model. Several attempts have been made in order to explain
such detections in the context of non-Gaussian scenarios or interacting dark energy models, showing that both these alternative 
cosmologies predict an enhanced number density of massive clusters at high redshifts, possibly alleviating the tension.
However, all the models proposed so far
also overpredict the abundance of massive clusters at the present epoch, and are therefore in contrast with
observational bounds on the low-redshift halo mass function. In this paper we present for the first time a new class of interacting dark energy
models that simultaneously account for an enhanced number density of massive clusters at high redshifts and for
both the standard cluster abundance at the present time and the standard power spectrum normalization at CMB. 
The key feature of this new class of models is the ``bounce" of the dark energy scalar field on the cosmological constant barrier at relatively recent epochs.
 We present the background and linear perturbations evolution of the model, showing that the standard amplitude of density perturbations 
 is recovered both at CMB and
 at the present time, and we demonstrate by means of large N-body simulations that our scenario predicts an enhanced number
 of massive clusters at high redshifts without affecting the present halo abundance. Such behavior could not arise in non-Gaussian
 models, and is therefore a characteristic feature of the bouncing coupled dark energy scenario.

\end{abstract}



\section{Introduction}
\label{i}

The origin of the observed acceleration of the cosmic expansion \citep{Riess_etal_1998,Perlmutter_etal_1999,SNLS,Kowalski_etal_2008} 
represents one of the most relevant open
questions in astrophysics and cosmology. Besides the standard $\Lambda $CDM cosmological model, that assumes
a perfectly homogeneous and static energy density represented by a cosmological constant $\Lambda $ as the source of the acceleration,
alternative scenarios based on a dynamical dark energy (DE) field such as {\em Quintessence} \citep[][]{Wetterich_1988,Ratra_Peebles_1988,
Ferreira_Joyce_1998} and {\em k-essence} \citep{ArmendarizPicon_etal_2000, kessence}, or on a modification of General Relativity
at large scales \citep[see \eg][]{Hu_Sawicki_2007,Appleby_Weller_2010} have been proposed in recent years. 

In order to efficiently distinguish among these different possibilities it is therefore
crucial to devise observational tests capable of highlighting a possible dynamical nature of the DE field, thereby falsifying the cosmological
constant picture. While present observational constraints on the background evolution of the Universe seem to be fully consistent
with a static behavior of the DE density, as expected for a standard cosmological constant, complementary and more sensitive tests
of a possible time evolution of the DE field might arise from the study of structure formation processes, and in particular from the time 
evolution of density perturbations at different scales.

In this context, the detection of unexpectedly massive clusters at high redshift \citep{Mullis_etal_2005, Bremer_etal_2006, Jee_etal_2009,Rosati_etal_2009,Brodwin_etal_2010, Jee_etal_2011,Foley_etal_2011} 
has recently attracted
significant interest as a possible indication of a failure of the standard $\Lambda $CDM cosmological model. In fact, starting from a
Gaussian field of primordial scalar perturbations with an amplitude consistent with the most recent constraints from observations of the
Cosmic Microwave Background (CMB) radiation \citep{wmap7}, 
the growth of density fluctuations for a $\Lambda $CDM cosmology predicts an evolution of the halo number density that would imply an extremely low probability
for the detection of such clusters. 
In particular, for the most extreme case represented by the XMMU J2235.3-2557 cluster \citep{Jee_etal_2009,Rosati_etal_2009}
with an estimated mass of $M_{324} = (6.4 \pm 1.2) \times 10^{14} M_{\odot}$ at $z\sim 1.4$, 
a $2-3\sigma $ discrepancy with respect to the fiducial $\Lambda $CDM model has been claimed by several authors \citep[see \eg][]{Jee_etal_2009,Jimenez_Verde_2009,Holz_Perlmutter_2010, Hoyle_Jimenez_Verde_2011, Jee_etal_2011},
although more conservative interpretations have been provided by other analysis \citep[as \eg by][]{Mortonson_Hu_Huterer_2011}. 
The main difficulties in driving sufficiently robust conclusions on a possible tension with the standard $\Lambda $CDM model
from such detections
arise as a consequence of the observational uncertainties on the mass determination of the clusters and on the estimation of the statistical
significance of the cosmological volumes covered by the cluster surveys \citep{Sheth_Diaferio_2011,Waizmann_Ettori_Moscardini_2011}. 
Such significance will need to be better clarified before claiming
a discrepancy with respect to the predictions of the standard model. Nevertheless, it is in any case interesting to investigate whether such possible
discrepancy could be alleviated by alternative cosmological scenarios.

Several attempts have been made in this direction in the recent past. On one side, a deviation
from statistical Gaussianity in the primordial density field has been shown to give rise to a higher number density of massive halos
as compared to the standard Gaussian case \citep[see \eg][]{Matarrese_Verde_Jimenez_2000, Grossi_etal_2007,Jimenez_Verde_2009,Holz_Perlmutter_2010,Cayon_Gordon_Silk_2010,Sartoris_etal_2010,Hoyle_Jimenez_Verde_2011,LoVerde_Smith_2010}.
However, in order to significantly increase the probability of detection this approach requires a level of primordial non-Gaussianity
that seems to be already ruled out by CMB constraints \citep{wmap7}, unless a strongly scale-dependent non-Gaussianity
is invoked \citep[as proposed by \eg][]{LoVerde_etal_2008,Cayon_Gordon_Silk_2010,Hoyle_Jimenez_Verde_2011}.
Alternatively, \citet{Baldi_Pettorino_2011} have shown that interacting DE models,
which are generically characterized by a faster growth of density perturbations 
with respect to $\Lambda $CDM
due to the presence of a fifth-force mediated by the DE scalar field, also predict a larger number density of massive halos at all redshifts 
that would result in a higher detection probability.

Both these approaches, however, are faced by the additional problem of overpredicting the abundance of massive clusters at low redshift.
In fact, while the detection of massive clusters at high redshifts seems to imply an anomalous growth of density perturbations as compared to
the predictions of the standard model,
the observed number counts at low redshifts are found to be in very good agreement with the predictions of $\Lambda $CDM \citep[see \eg][]{Reiprich_Boehringer_2002,
Vikhlinin_etal_2009b, Mantz_etal_2010, Rozo_etal_2010}.
In other words, a non-Gaussian initial density field or a faster growth of density fluctuations can determine an increased number density of massive
halos at high redshifts, thereby alleviating the possible tension with the model arising from the unexpected detection of an exceeding number of such objects, but will then 
also necessarily imply the further growth of such massive halos from these high redshifts to the present time, resulting in a 
clear discrepancy with present bounds on the cluster mass function at low redshifts.

In the present study we will show how this problem can be naturally solved by an interacting DE model with a suitable self-interaction potential different from the standard 
exponential or power-law potentials assumed in previous investigations of coupled DE (cDE) cosmologies. 
In particular, we will show how the existence of a global minimum in the potential, and the consequent ``bounce" of the DE field on the cosmological constant barrier, 
is the key feature to address the existence of 
exceedingly massive clusters at high redshift without necessarily affecting the halo mass function at the present epoch.
\ \\

Although our proposed scenario does not benefit in general of the same scaling properties typical of the exponential and power-law potentials, 
it  naturally provides a 
mechanism to enhance the expected number of massive halos up to some high redshift and to subsequently reduce it again to
the standard $\Lambda $CDM prediction at the present time, thereby accounting at the same time for the existence of massive clusters at early epochs
and for the observed halo abundance at low redshift.
We therefore regard the specific model presented in this work as a toy example of how the dynamical nature of a DE scalar field interacting with cold dark matter (CDM)
particles can account for the presence of a high redshift peak in the deviation of the expected cluster number counts from the $\Lambda $CDM
prediction. This feature could not arise in standard gravity models or in non-Gaussian cosmological scenarios, and would therefore represent a ``smoking gun"
for the dynamical nature of DE.
In the present work we will illustrate in detail this peculiar behavior, also by means of large N-body simulations of structure formation.
\ \\

The paper is organized as follows. In Section~\ref{sec:cDE} we briefly review the main features of standard cDE models
for the background evolution of the universe, and we highlight how bouncing cDE models deviate from this standard and well-known behavior.
In Section~\ref{sec:linear} we investigate the evolution of linear density perturbations in the context of both standard and bouncing cDE scenarios,
and we show how the latter can provide a possible explanation for a larger amplitude of density perturbations at high redshift without affecting
the normalization of the linear matter power spectrum at the present time. This effect is further investigated for the nonlinear regime of structure formation 
in Section~\ref{sec:sims},
where we show by means of large N-body simulations how the real halo
mass function evolves in time in each of the models considered, and how bouncing cDE models can predict an enhanced
number of massive clusters at high redshift without changing the halo abundance at the present epoch. 
Finally, in Section~\ref{sec:concl} we draw our conclusions.

\section{Interacting dark energy models with exponential and SUGRA potentials}
\label{sec:cDE}
\begin{figure*}
\includegraphics[scale=0.45]{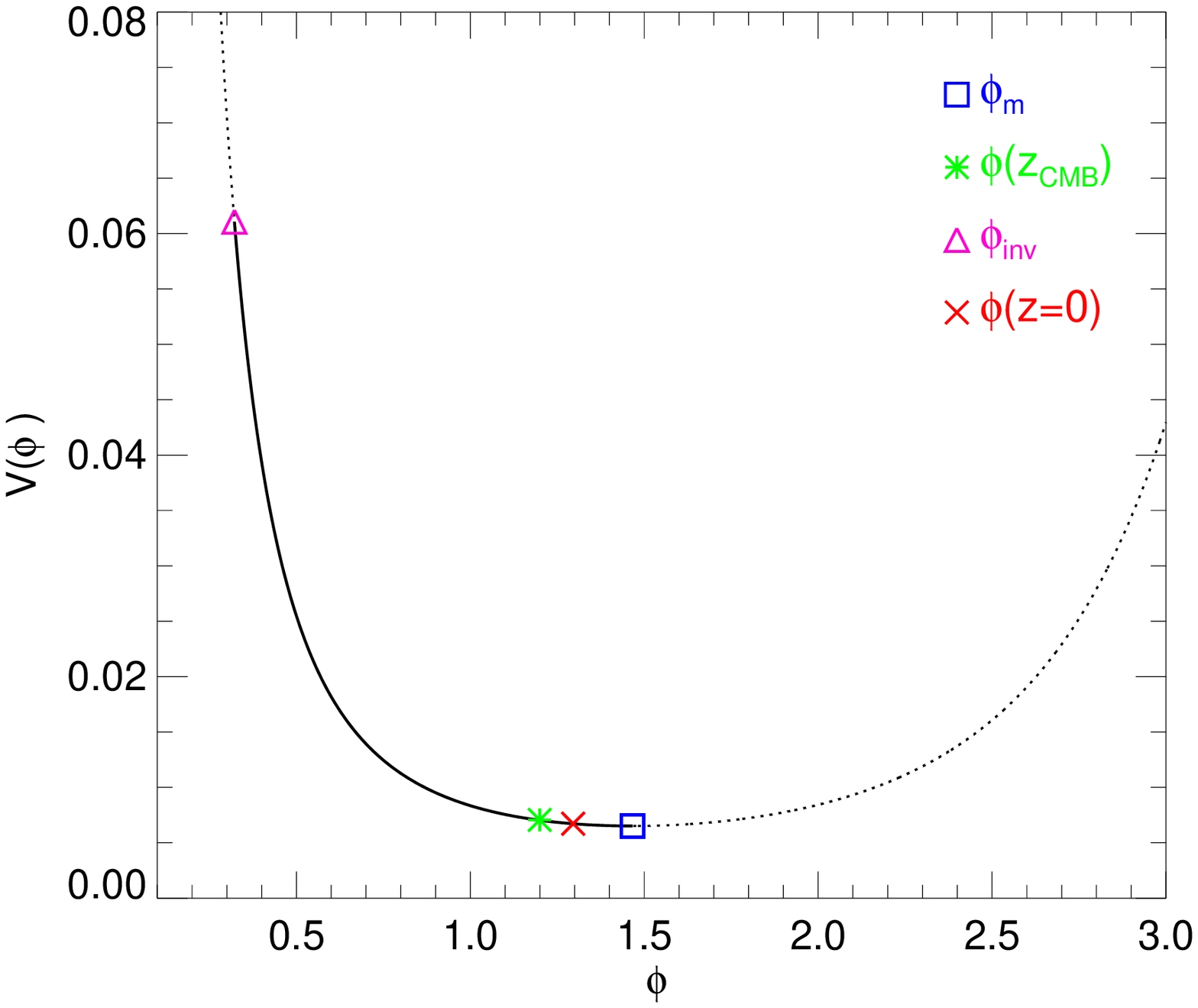}
\includegraphics[scale=0.45]{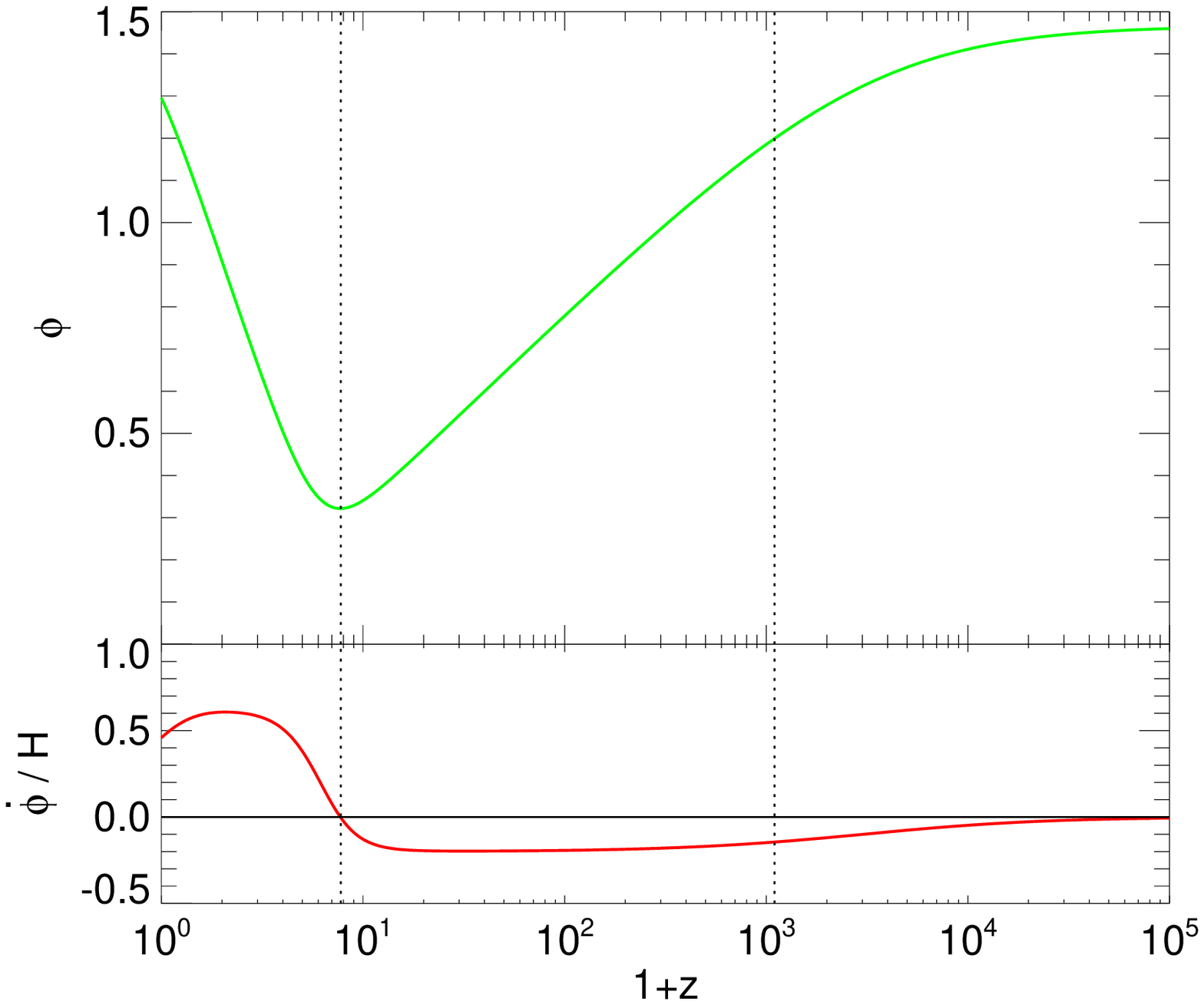}
\caption{{\em Left:} The trajectory of the DE scalar field $\phi $ along its self-interaction SUGRA potential during the whole expansion history of the universe.
The field is set at rest in the minimum of the potential at $z\rightarrow \infty$ (blue square) and subsequently moves away from the minimum
due to the effective force exerted by the coupling term in the Klein-Gordon equation (\ref{klein_gordon}). The field reaches the position marked by the green 
asterisk at $z_{\rm CMB}$ and keeps moving up the potential until the inversion point at $z_{\rm inv}\sim 6.8$ (pink triangle) where it inverts
its motion and rolls back down towards the minimum, reaching at $z=0$ the position marked by the red cross. {\em Right:} 
The same dynamical evolution described in the left panel: the upper plot shows the evolution in redshift of the scalar field $\phi $ while the 
lower plot displays the corresponding evolution of the specific scalar velocity $\dot{\phi }/H.$}
\label{fig:scalar_field}
\end{figure*}
\normalsize

Interacting DE models involving a direct exchange of energy-momentum between a DE scalar field  $\phi $
and other matter components of the universe have been widely investigated in the past, in particular 
for the case of a coupling with the CDM fluid \citep{Wetterich_1995,Amendola_2000,Amendola_2004,Pettorino_Baccigalupi_2008,Baldi_2011a}
or with massive neutrinos \citep{Amendola_Baldi_Wetterich_2008,Baldi_etal_2011a}. 
The effects of these interactions on observable quantities such as the CMB angular power spectrum \citep{Bean_etal_2008,Xia_2009,LaVacca_etal_2009}, 
the Lyman-$\alpha $ forest \citep{Baldi_Viel_2010}, the large scale structure of the universe \citep{Baldi_Pettorino_2011}, or the structural properties of highly nonlinear objects 
\citep{Baldi_etal_2010,Li_Barrow_2011,Baldi_2011b,Baldi_Lee_Maccio_2011} have been extensively studied by means of  both analytic and numerical techniques.

Most of the previous works on interacting DE models assume a monotonic function for the self-interaction potential of the DE scalar field,
as \eg an exponential \citep{Wetterich_1988} or an inverse power law \citep{Ratra_Peebles_1988} potential, which determine the existence
of attractor solutions that are almost independent on the initial conditions of the scalar field. These potentials do not possess local minima,
and therefore define a unique direction for the time evolution of the scalar field $\phi $. 
However, other possible functions, characterized in some cases 
by the presence of local or global minima, have been considered in the literature for the case of non-interacting
DE scenarios, as \eg the SUGRA potential \citep{Brax_Martin_1999} arising naturally in
supersymmetric theories of gravity. The existence of a local or a global minimum in the scalar field self-interaction potential allows the field
to oscillate and invert its motion during the expansion history of the Universe, as recently suggested by \citet{Tarrant_etal_2011}.
Such feature could have very significant consequences if a coupling to CDM is also present, as we will show in the 
present work.

We will study the case of an interacting DE model with a SUGRA self-interaction potential
and compare it with the standard $\Lambda $CDM cosmology and with coupled DE models with an exponential potential,
with a particular focus on the different effects that these two types of potential induce on the evolution of linear and nonlinear 
density perturbations in the universe and on the predicted number density of massive clusters at different cosmic epochs.
\ \\

We consider a flat cosmological model described by the following set of dynamic equations:
\begin{eqnarray}
\label{klein_gordon}
\ddot{\phi } + 3H\dot{\phi } +\frac{dV}{d\phi } &=& \sqrt{\frac{2}{3}}\beta _{c} \frac{\rho _{c}}{M_{Pl}} \,, \\
\label{continuity_cdm}
\dot{\rho }_{c} + 3H\rho _{c} &=& -\sqrt{\frac{2}{3}}\beta _{c}\frac{\rho _{c}\dot{\phi }}{M_{Pl}} \,, \\
\label{continuity_baryons}
\dot{\rho }_{b} + 3H\rho _{b} &=& 0 \,, \\
\label{continuity_radiation}
\dot{\rho }_{r} + 4H\rho _{r} &=& 0\,, \\
\label{friedmann}
3H^{2} &=& \frac{1}{M_{Pl}^{2}}\left( \rho _{r} + \rho _{c} + \rho _{b} + \rho _{\phi} \right)\,,
\end{eqnarray}
where the role of DE is played by the classical scalar field $\phi $ and where the subscripts $b$, $c$, and $r$ represent the baryonic, CDM and 
relativistic fluids, respectively. In Eqs.~(\ref{klein_gordon}-\ref{friedmann}) an overdot represents a derivative with respect to the cosmic time $t$, $H\equiv \dot{a}/a$, and $M_{Pl}\equiv 1/\sqrt{8\pi G}$ is the reduced Planck mass
\footnote{Notice that in Eqs.~(\ref{klein_gordon}-\ref{friedmann}) and throughout the rest of this paper we assume the same convention on the definition of the coupling adopted by \citet{Amendola_2000} and \citet{Baldi_2011a}, which differs by a factor $\sqrt{3/2}$ from the definition adopted in some of the literature.}.
In this work we will restrict our analysis to the case of a constant coupling $\beta _{c}$, although a direct dependence of the coupling on the scalar field
could also be considered \citep[see \eg][]{Baldi_2011a}.

As already anticipated, besides the standard $\Lambda $CDM model we will consider two possible choices for the scalar field self-interaction potential $V(\phi )$,
namely an exponential potential \citep{Lucchin_Matarrese_1984,Wetterich_1988,Ferreira_Joyce_1998} and a SUGRA potential \citep{Brax_Martin_1999}, 
defined as:
\begin{eqnarray}
{\rm EXP:}\quad V(\phi ) &=& Ae^{-\alpha \phi } \,,\\
{\rm SUGRA:}\quad V(\phi ) &=& A\phi ^{-\alpha }e^{\phi ^{2}/2}\,,
\end{eqnarray}
where the scalar field $\phi $ is normalized in units of the reduced Planck mass. 
For the exponential models we will assume a positive coupling with values $\beta _{c}=0.1$ and $\beta _{c} = 0.15$, 
while for the SUGRA model
a negative coupling $\beta _{c} = -0.15$.
Our proposed SUGRA cDE scenario therefore requires the same number of parameters as standard cDE models.

We compute the background evolution of these models by numerically integrating
the system of Eqs.~(\ref{klein_gordon}-\ref{friedmann}) with a trial-and-error procedure until
the the desired values of the background cosmological parameters at $z=0$ are obtained.
These correspond to the ``WMAP7-only Maximum Likelihood"
parameters derived by \citet{wmap7}, and are listed in Table~\ref{tab:parameters}.
\begin{table}
\begin{center}
\begin{tabular}{cc}
\hline
Parameter & Value\\
\hline
$H_{0}$ & 70.3 km s$^{-1}$ Mpc$^{-1}$\\
$\Omega _{\rm CDM} $ & 0.226 \\
$\Omega _{\rm DE} $ & 0.729 \\
$\sigma_{8}$ & 0.809\\
$ \Omega _{b} $ & 0.0451 \\
$n_{s}$ & 0.966\\
\hline
\end{tabular}
\end{center}
\caption{The cosmological parameters at $z=0$ assumed for all the models considered in the present study, consistent with the
``WMAP7 only Maximum Likelihood" results of \citet{wmap7}.}
\label{tab:parameters}
\end{table}
All the models considered in the present work with
the relative parameters and normalizations are instead summarized in Table~\ref{tab:models}.
\begin{table*}
\begin{tabular}{llcccccccc}
Model & Potential  &  
$\alpha $ &
$\beta _{0}$ &
$\beta _{1}$ &
\begin{minipage}{45pt}
Scalar field \\ normalization
\end{minipage} &
\begin{minipage}{45pt}
Potential \\ normalization
\end{minipage} &
$w_{\phi }(z=0)$ &
${\cal A}_{s}(z_{\rm CMB})$ &
$\sigma _{8}(z=0)$\\
\\
\hline
$\Lambda $CDM & $V(\phi ) = A$ & -- & -- & -- & -- & $A = 0.0219$ & $-1.0$ & $2.42 \times 10^{-9}$ & $0.809$ \\
EXP002 & $V(\phi ) = Ae^{-\alpha \phi }$  & 0.08 & 0.1 & 0 &$\phi (z=0) = 0$ & $A=0.0218$ & $-0.995$ & $2.42 \times 10^{-9}$ & $0.875$ \\
EXP003 & $V(\phi ) = Ae^{-\alpha \phi }$  & 0.08 & 0.15 & 0 & $\phi (z=0) = 0$ & $A=0.0218$ & $-0.992$ & $2.42 \times 10^{-9}$ & $0.967$\\
SUGRA003 & $V(\phi ) = A\phi ^{-\alpha }e^{\phi ^{2}/2}$  & 2.15 & -0.15 & 0 & $\phi (z\rightarrow \infty ) = \sqrt{\alpha }$ & $A=0.0202$ & $-0.901$ & $2.42 \times 10^{-9}$ & $0.806$ \\
\hline
\end{tabular}
\caption{The defining parameters and their normalization for the different cosmological models considered in the present work. All the models are consistent with the same amplitude of scalar perturbations ${\cal A}_{s}$ at $z_{\rm CMB} \approx 1100$, but due to the different growth of perturbations the amplitude of the linear
matter power spectrum at $z=0$ represented by $\sigma _{8}$ is different from model to model. Notice that the SUGRA cDE cosmology is the only model to have the same amplitude of linear
perturbations as the standard $\Lambda $CDM model both at $z_{\rm CMB}$ and at $z=0$.}
\label{tab:models}
\end{table*}
\ \\

Since the main features of the  cDE models with an exponential potential have been extensively discussed in the literature \citep[see \eg][]{Wetterich_1995,Amendola_2000,Amendola_2004,
Pettorino_Baccigalupi_2008,Baldi_2011a,Baldi_2011b}
we refer the interested reader to these publications and we will mainly focus our attention here on the dynamics of the new proposed
SUGRA cDE scenario which has never been presented before.
The key feature of the SUGRA potential for the analysis carried out in this work is the presence of a global minimum at $\phi _{\rm m} = \sqrt{\alpha }$,
where the derivative of the potential in Eq.~(\ref{klein_gordon}) vanishes. 
For our model, we assume the scalar field $\phi $ to be at rest in the global minimum at very early times. This is a natural choice
for the initial conditions since any previous dynamical evolution of the scalar field could be efficiently damped by Hubble friction in the radiation
dominated epoch.

In the absence of any coupling to CDM the global minimum of the SUGRA potential would be a stable critical point of the system, 
and the scalar field would therefore remain at rest at its initial location $\phi _{\rm m}$ during the whole expansion history of the universe, effectively behaving
as a cosmological constant. 
On the contrary, if a coupling to CDM is present, the Klein-Gordon equation (\ref{klein_gordon}) for the scalar field at rest in the minimum reads:
\begin{equation}
\phi _{i}=\phi _{\rm m}\,, \, \, \dot{\phi }_{i}=0 \quad \Rightarrow \quad \ddot{\phi } = \sqrt{\frac{2}{3}}\beta _{c} \frac{\rho _{c}}{M_{Pl}} \,,
\end{equation}
showing that a non-vanishing acceleration term due to the coupling acts on the scalar field as soon as the CDM density $\rho _{c}$
starts to play a significant role in the dynamical evolution of the universe. In particular, for our choice of the parameters in the SUGRA cDE model
the negative coupling $\beta _{c}=-0.15$ will push the field away from its initial location $\phi _{\rm m}$ in the direction of smaller
field values, \ie towards the power-law side of the SUGRA potential. As soon as the scalar field $\phi $ leaves the local minimum with a 
negative velocity ($\dot{\phi }<0$) the other terms appearing in Eq.~(\ref{klein_gordon}) are restored and start to slow down the field,
which will eventually stop and invert its motion moving again towards the minimum of the SUGRA potential.

The details of this peculiar dynamic evolution depend crucially on the balance between the different parameters of the model, and in particular
on the ratio between the slope of the SUGRA potential $\alpha $ and the coupling strength $\beta _{c}$. It is therefore necessary to tune 
these parameters quite accurately in order to obtain a viable solution for the background evolution of the universe, and the model therefore 
presents a comparable level of fine tuning as for the standard $\Lambda $CDM cosmology. Nevertheless, as we will describe below, this scenario
presents several appealing features and provides a self consistent way to address the issue of anomalous massive clusters at high redshift.
It should therefore be considered as a toy example of the possible characteristic signatures of a dynamic DE field on structure formation.

In the two panels of Fig.~\ref{fig:scalar_field} we show the dynamic evolution of the scalar field $\phi $ in the context of our proposed SUGRA cDE model. In the left
panel we plot the evolution of the field along the SUGRA self-interaction potential at different times. The dotted curve shows the SUGRA
potential for our choice of parameters while the solid line shows the trajectory of the field along the potential. 
The field is set to be at rest in the minimum of the 
potential $\phi _{\rm m} = \sqrt{\alpha } \approx 1.47$ at early times (as shown by the blue square) and then evolves towards smaller field values
reaching the position indicated by the green asterisk at the redshift of CMB, $z_{\rm CMB}\sim 1100$. The location of the scalar field at $z_{\rm CMB}$
is particularly relevant for the analysis carried out in the present work, as will be extensively discussed below. 
After $z_{\rm CMB}$ the field keeps then moving up the potential
and eventually stops and inverts its motion at redshift $z_{\rm inv}$ (shown in Fig.~\ref{fig:scalar_field} by the pink triangle) which for the specific model considered here is at $z_{\rm inv} \sim 6.8$, 
and finally rolls back down towards the minimum, reaching at $z=0$ the position indicated by the red cross. It is important to notice here, for reasons that will be clarified later on,  
that the field values at $z_{\rm CMB}$ and at $z=0$ are relatively close to each other. 

The same dynamics are displayed in the right panel of Fig.~\ref{fig:scalar_field},
where the evolution of the scalar field $\phi $ (upper plot) and of its specific velocity $\dot{\phi }/H$ (lower plot) are shown as a function of redshift.
Once more, the plot shows how the field starts at rest in the minimum of the SUGRA potential 
at high redshifts and then acquires a negative velocity due to the negative coupling term until the motion is eventually inverted 
and its velocity becomes positive at $z_{\rm inv}$. The vertical dotted lines indicate the two redshifts $z_{\rm CMB}$ and $z_{\rm inv}$.

\begin{figure}
\includegraphics[scale=0.45]{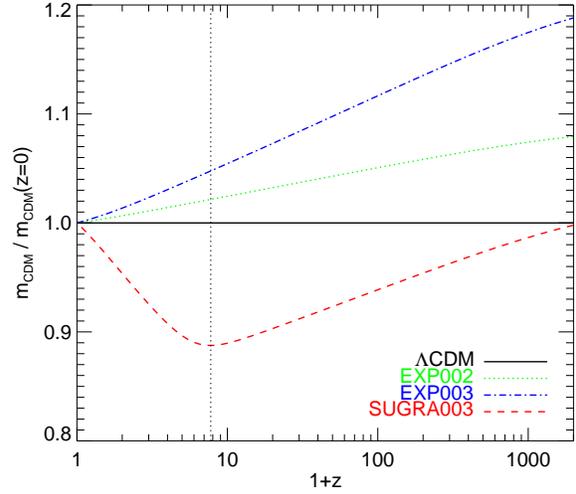}
\caption{The time evolution of the CDM particle mass in the different cDE models. The different curves show the CDM particle mass in units of its present-day value.
While the standard exponential cDE models show a monotonic behavior of the CDM mass, with a continuous flux of energy from CDM to DE, the 
SUGRA cDE model, due to the ``bounce" of the scalar field and the inversion of its motion features a phase of CDM mass decrease followed by a
subsequent increase. The trend is inverted in correspondence of the scalar field bounce, indicated in the figure by the vertical dotted line.}
\label{fig:mass}
\end{figure}
\normalsize

The inversion of the motion of the scalar field during cosmic evolution in our SUGRA cDE scenario, as opposed to standard
cDE models with monotonic potentials where the scalar field moves always in the same direction for the whole expansion
history of the universe, plays the crucial role in determining the effects on the growth of density perturbations that we will discuss in
the next Section. In fact, as it is well known \citep[see \eg][]{Amendola_2000,Baldi_etal_2010,Tarrant_etal_2011}, 
the dynamic evolution of the DE scalar filed in the context of cDE models determines a 
transfer of energy-momentum between the DE and CDM fluids, according to the strength and sign of the coupling constant $\beta _{c}$.
As one can clearly see from the CDM continuity equation (Eq.~\ref{continuity_cdm}), the transfer of energy between DE and CDM is directly
proportional to the product of the coupling constant $\beta _{c}$ and of the scalar field velocity $\dot{\phi }$. With our chosen convention, a positive
combination $\beta _{c}\dot{\phi } > 0$ implies a flux of energy from the CDM to the DE component, while a negative combination $\beta _{c}\dot{\phi } < 0$
implies the opposite direction for the energy transfer. By integrating Eq.~(\ref{continuity_cdm}) and by assuming the conservation of the total CDM particle number
one can therefore compute the time variation of the mass of CDM particles as:
\begin{equation}
\label{mass_variation}
m_{c}(z) = m_{c,0}\cdot e^{-\beta _{c}\phi (z)} \,,
\end{equation}
where $m_{c,0}$ denotes the value of the CDM particle mass at $z=0$.
It is therefore clear from Eq.~(\ref{mass_variation}) that for the case of cDE models with a monotonic potential the CDM particle mass will be 
always decreasing or always increasing, while for the SUGRA cDE scenario the time derivative of the CDM particle mass $\dot{m}_{c}$ will
change sign during cosmic evolution in correspondence to the inversion of motion of the DE scalar field, at $z_{\rm inv}$.
A similar behavior has been recently described also in the context of Growing Neutrino quintessence models \citep{Baldi_etal_2011a}.
This is shown in Fig.~\ref{fig:mass}, where the evolution of the CDM particle mass for the $\Lambda $CDM, EXP002, EXP003, and SUGRA003 models 
(solid black, dotted green, dot-dashed blue, and dashed red lines, respectively) is plotted as a function of redshift. As one can see from the plot,
while the mass is always decreasing in the standard EXP002 and EXP003 cDE models, indicating a continuous flux of energy from the CDM to the DE, the mass
in the SUGRA003 model first decreases (when $\dot{\phi } < 0$) and then increases again (when $\dot{\phi } > 0$), such that the CDM particle mass
at $z_{CMB}$ has almost the same value it has at $z=0$,  which is clearly not the case for a standard cDE scenario with a monotonic potential. 
This is obviously a consequence of the fact that for our specific choice of parameters and initial conditions
the final location of the scalar field is relatively close, as mentioned above, to its location at the time of CMB.
Such peculiar evolution of the CDM particle mass will have a significant impact on the growth of density perturbations, as we will discuss in the next Section.
\begin{figure*}
\includegraphics[scale=0.45]{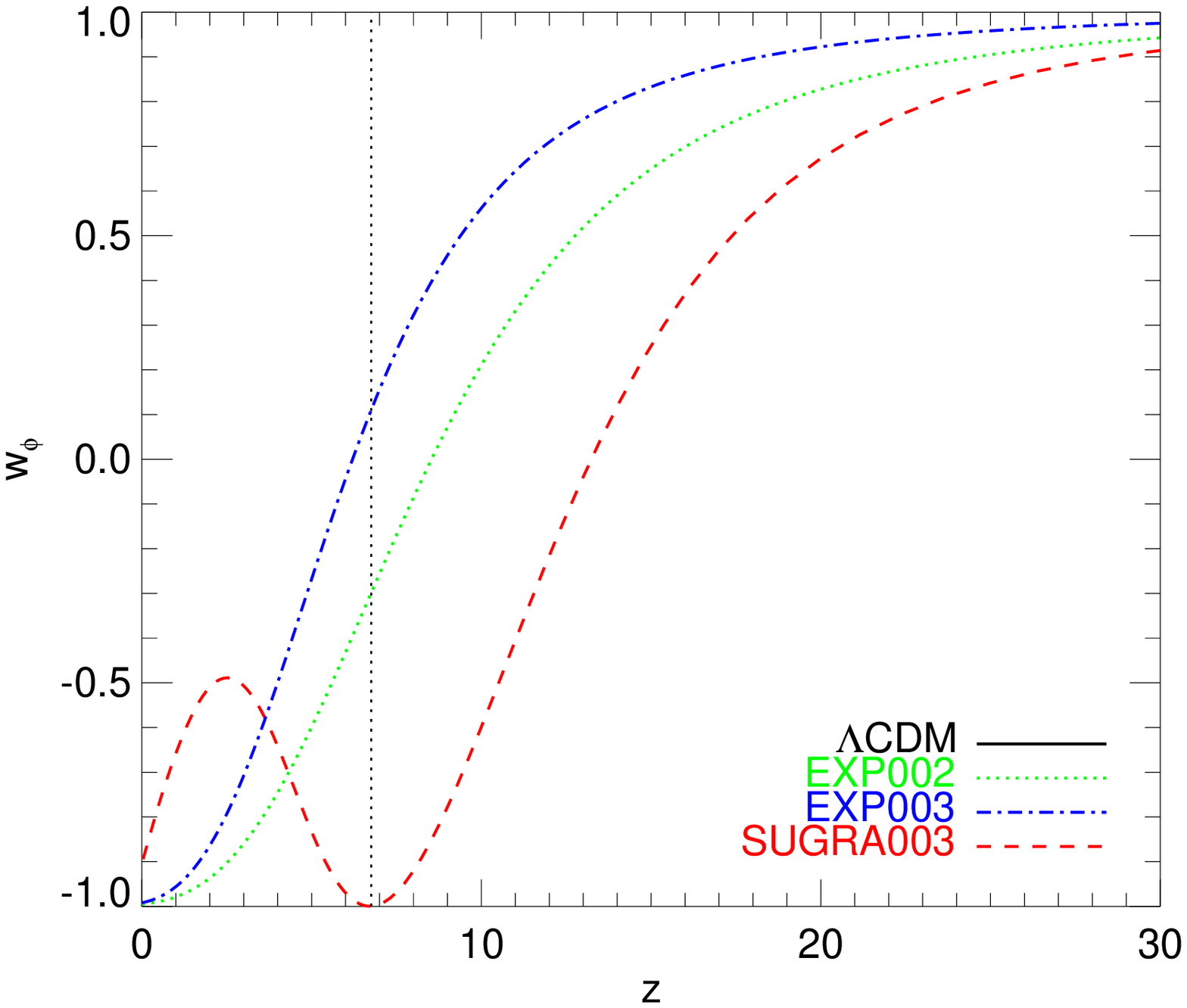}
\includegraphics[scale=0.45]{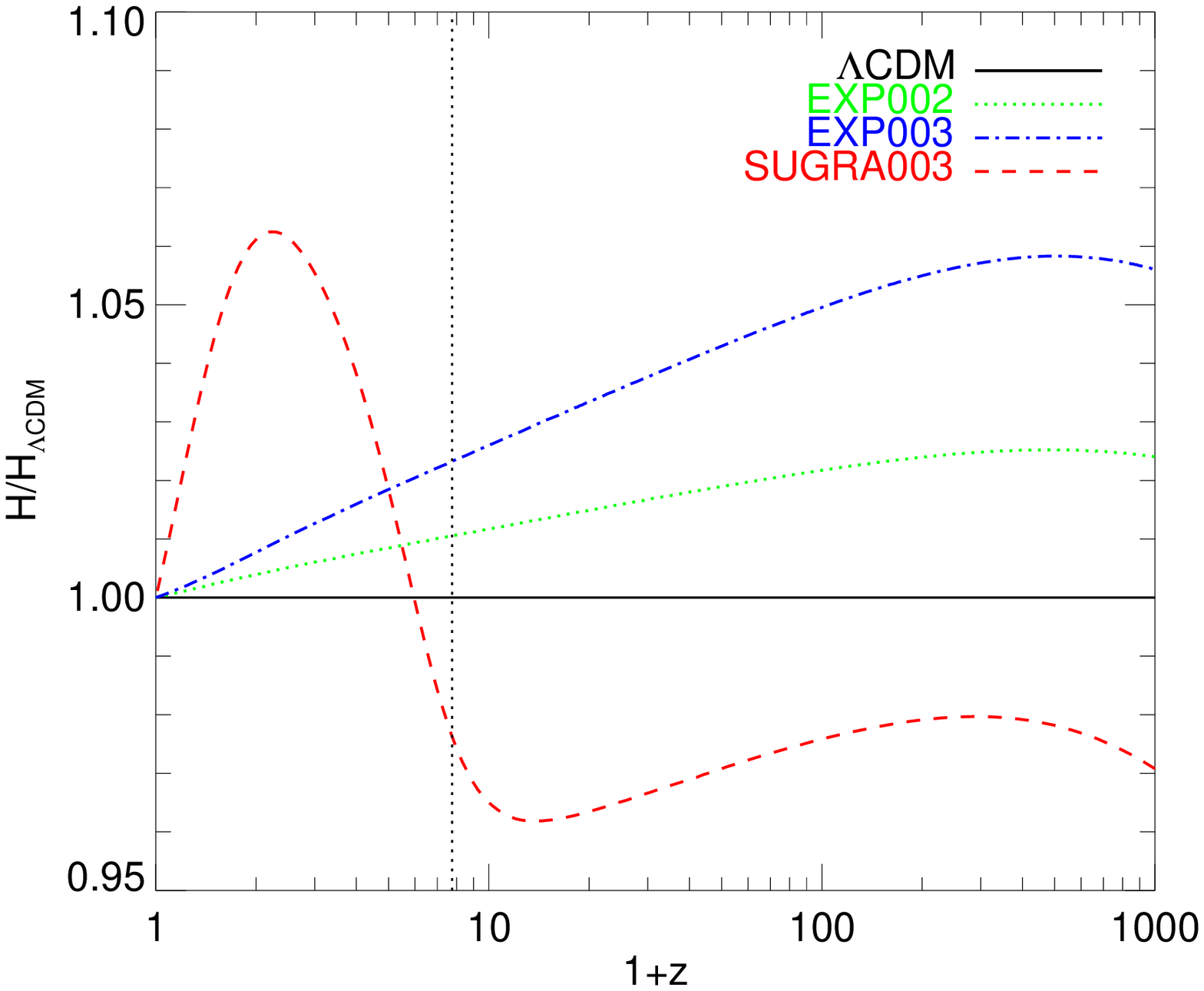}
\caption{{\em Left:} The evolution of the DE equation of state parameter $w_{\phi }$ as a function of redshift.
While the standard exponential cDE models show a monotonic evolution of $w_{\phi }$ that asymptotically approaches
the cosmological constant value of $-1$ at low redshifts, the SUGRA cDE model features a ``bounce" of the equation of state parameter
on the cosmological constant barrier (sometimes also called the {\em phantom divide}) at $z_{\rm inv}\sim 6.8$, before decreasing again towards
$-1$ at lower redshifts, reaching at the present time a value of $w_{\phi }(z=0) = -0.901$.
{\em Right:} The ratio of the Hubble function $H(z)$ over
the standard $\Lambda $CDM case as a function of redshift. The non-trivial dynamics at low redshifts of the DE scalar field $\phi $ in the SUGRA cDE
model imprints a specific signature on the Hubble function. For this model the deviation from $\Lambda $CDM is larger at very low redshifts as compared to
standard exponential cDE scenarios, reaching a deviation of $\sim 6\%$ at $z=1$, while at very high redshifts the deviation from $\Lambda $CDM
is smaller with respect to a standard cDE model with identical coupling strength (EXP003).}
\label{fig:background}
\end{figure*}
\normalsize
\ \\

Before moving to consider the formation of cosmic structures in the context of our proposed SUGRA cDE model, we conclude the analysis of the background evolution
by computing the Hubble function and the DE equation of state parameter, defined as $w_{\phi }\equiv (\dot{\phi }^{2}/2 - V(\phi))/(\dot{\phi }^{2}/2 + V(\phi ))$, for the 
four models considered in our comparison. In Fig.~\ref{fig:background} we show in the left panel the evolution of $w_{\phi }$ with redshift for the EXP002, EXP003
and SUGRA003 models. As one can clearly see, the bounce of the scalar field at $z_{\rm inv}$ in the SUGRA cDE model, where the scalar field velocity
$\dot{\phi }$ changes sign, is reflected in a bounce of the DE equation of state on the cosmological constant barrier at $w_{\phi } = -1$. The following 
acceleration of the scalar field towards the minimum of the potential induces therefore a temporary increase of $w_{\phi }$ up to $w_{\phi }\sim -0.5$ at $z\sim 2.5$
until the Hubble friction starts decelerating again the scalar field motion, which ends up with an equation of state parameter $w_{\phi }=-0.901$ at $z=0$.
This non-trivial evolution of the equation of state parameter is not present for the standard exponential cDE models where $w_{\phi }$ asymptotically approaches
$-1$ at low redshifts, and could therefore possibly strongly constrain the SUGRA cDE scenario. Its impact on the Hubble expansion, in fact, is quite significant at 
relatively low redshifts which could be probed by constraints on the expansion history coming from Supernovae Ia or from
accurate measurements of the Baryon Acoustic Oscillations scale \citep[see \eg][]{Simpson_Bridle_2006} from present and future wide surveys, 
This is shown in the right panel of Fig.~\ref{fig:background}, where the ratio of the Hubble function for the EXP002, EXP003 and SUGRA003 cDE
models (dotted green, dot-dashed blue and dashed red lines, respectively) to the standard $\Lambda $CDM case is plotted as a function of redshift.
As one can see from the plot, the SUGRA cDE model determines a stronger difference from the standard $\Lambda $CDM Hubble function with respect to an exponential
cDE model with the same coupling strength at low redshifts, with a maximum deviation of about $6\%$ at $z\sim 1$, while at high redshifts the SUGRA model is
found to be closer to $\Lambda $CDM as compared to the exponential case. 

This low-redshift feature of the SUGRA cDE might challenge our model by
putting it in tension with observational constraints on the local expansion
history.
Nevertheless, a significant dynamical evolution of the DE scalar field at $z<2$ is a necessary condition in order to account for the existence of anomalously massive
objects at high redshifts and simultaneously recover the $\Lambda $CDM amplitude of density perturbations at $z=0$, as we will discuss in full detail in Section~\ref{sec:linear}.
Furthermore, it is important to stress here that we are considering only one specific and somewhat extreme realization of the SUGRA cDE scenario, 
in order to enhance the effects under investigation, while a detailed parameter fitting
would be required to select the set of parameters that best reproduce presently available background observations. Ultimately, also the specific form of
the self-interaction potential could be changed, since for our arguments the choice of the SUGRA potential is motivated mainly by its firm theoretical origin
in the context of supersymmetric theories of gravity, while also other functional forms of the potential could work equally well, provided they feature a global minimum.
We are therefore proposing this specific model as a toy example of how a global or a local minimum in the DE self-interaction potential could 
provide a mechanism to explain unexpected detections of massive clusters at high redshift, while a detailed survey of such models would be required in 
order to assess their viability for the background evolution of the universe.

\section{Linear perturbations}
\label{sec:linear}

The evolution of linear density perturbations in the context of cDE cosmologies has been extensively studied by several authors 
\citep[see \eg][]{Amendola_2004,Pettorino_Baccigalupi_2008,Baldi_etal_2010,Baldi_2011a} to which we refer for the derivation of the 
main perturbations equations in the presence of a coupling between DE and CDM.
In the present Section we recall the main modifications introduced by the coupling on the growth of density perturbations
in the linear regime and we highlight how the dynamical evolution of a SUGRA cDE scenario extensively described above
can substantially alter this process and introduce noticeable features in the CDM growth factor.

As a starting point, we consider the evolution equation for CDM and baryon density perturbations $\delta _{c,b}\equiv \delta \rho _{c,b}/\rho _{c,b}$
in the Newtonian limit and in the presence of a DE-CDM interaction as derived by \eg \citet{Amendola_2004}, in Fourier space and in cosmic time:
\begin{eqnarray}
\label{gf_c}
\ddot{\delta }_{c} &=& -2H\left[ 1 - \beta _{c}\frac{\dot{\phi }}{H\sqrt{6}}\right] \dot{\delta }_{c} + 4\pi G \left[ \rho _{b}\delta _{b} + \rho _{c}\delta _{c}\Gamma _{c}\right] \,, \\
\label{gf_b}
\ddot{\delta }_{b} &=& - 2H \dot{\delta }_{b} + 4\pi G \left[ \rho _{b}\delta _{b} + \rho _{c}\delta _{c}\right]\,.
\end{eqnarray}
The additional contribution appearing in the first term on the right hand side of Eq.~(\ref{gf_c}) is the extra friction associated with
momentum conservation in cDE models \citep[see \eg ][for a discussion on the effects of the friction term]{Baldi_2011b} and
the factor $\Gamma _{c}$ defined as
\begin{equation}
\label{Gamma_c_massless}
\Gamma _{c} = 1 + \frac{4}{3} \beta ^{2}_{c}
\end{equation}
includes the effect of the fifth-force mediated by the DE scalar field for CDM density perturbations.

A few assumptions have to be made in order to derive Eqs.~(\ref{gf_c},\ref{gf_b}), besides the already mentioned 
Newtonian limit of small scales for the Fourier modes under consideration, \ie $\lambda \equiv aH/k \ll 1$.
In particular, one has to assume the dimensionless mass of the scalar field $\phi $, defined as 
\begin{equation}
\hat{m}^{2}_{\phi} \equiv \frac{1}{H}\frac{d^{2}V}{d\phi ^{2}} \,,
\end{equation}
to be small compared to the Fourier modes at the scales of interest, such that
\begin{equation}
\hat{m}^{2}_{\phi }\lambda ^{2}\ll 1\,.
\end{equation}
If this condition is not fulfilled, the clustering of the DE scalar field $\phi $ might grow beyond the linear level at scales below $1/\hat{m}^{2}_{\phi }$
and the fifth force of Eq.~(\ref{Gamma_c_massless}) would then acquire a Yukawa suppression factor given by:
\begin{equation}
\label{Gamma_c_massive}
\Gamma _{c} \equiv 1 + \frac{4}{3}\frac{\beta _{c}^{2}}{1 + \hat{m}^{2}_{\phi }\lambda ^{2}} \,.
\end{equation}
Since both the linear treatment of density perturbations discussed in the present Section and the non-linear N-body algorithm used
for the analysis presented in the next Section are based on the assumption of a small scalar mass, and therefore on a term
like (\ref{Gamma_c_massless}) for the fifth-force implementation, it is important to clarify to which extent this assumption can be considered
to hold. 
\begin{figure}
\includegraphics[scale=0.45]{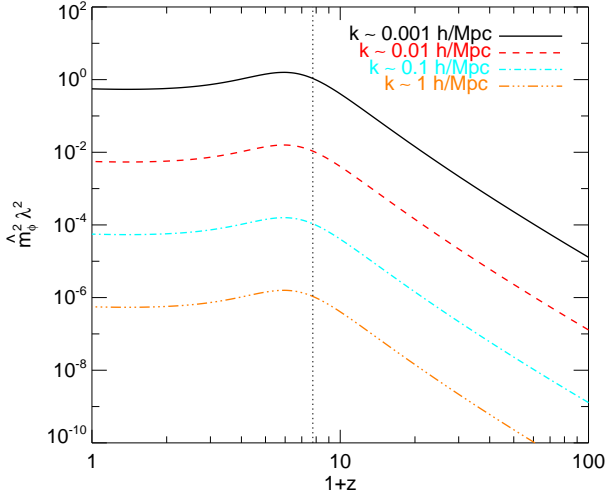}
\caption{The time evolution of the quantity $\hat{m}^{2}_{\phi }\lambda ^{2}$ defining the validity of the linear approximation adopted both for the integration
of the linear growth factor and for the N-body implementation used in the present work. As the Figure shows, $\hat{m}^{2}_{\phi }\lambda ^{2}$ gets close to unity
only at extremely large scales ($k\sim 0.001\, k/$Mpc). At the scales of interest in this work, the linear approximation assumed to derive Eqs.~(\ref{gf_c},\ref{gf_b}) is therefore fully
justified.}
\label{fig:scalar_mass}
\end{figure}
\normalsize
In Fig.~\ref{fig:scalar_mass} we plot the quantity $\hat{m}^{2}_{\phi }\lambda ^{2}$ for several comoving wavenumbers $k\sim 1\,,0.1\,,0.01\,,0.001\, h/$Mpc
as a function of redshift. As one can see from the plot, the scalar mass can start playing a significant role at $z<10$ only for scales close to the cosmic horizon,
while for all scales below $\sim 700$ comoving Mpc$/h$ ($k\sim 0.01\, h/$Mpc) the influence of a non-zero scalar mass is negligible and Eqs.~(\ref{gf_c},\ref{gf_b},\ref{Gamma_c_massless}) can be safely used. For the N-body results discussed in Sec.~\ref{sec:sims} we will consider simulations with a box
of $1$ comoving Gpc$/h$ aside, for which only the largest scales sampled by the initial power spectrum  might be marginally affected by our massless field approximation,
while we will concentrate on nonlinear structure formation processes occurring at much smaller scales.

Having clarified the range of validity of the assumptions made in deriving Eqs.~(\ref{gf_c},\ref{gf_b},\ref{Gamma_c_massless}) we can now discuss one of the
central results of the present work. By numerically solving the system of Eqs.~(\ref{klein_gordon}-\ref{friedmann},\ref{gf_c},\ref{gf_b}) at subhorizon scales
($k\sim 0.1\, h$/Mpc) we can compute the linear growth of density perturbations for all the different cosmological models under investigation and compare it to the standard
$\Lambda $CDM growth factor. As boundary conditions for our integration we impose that the ratio of baryonic to CDM perturbations at $z_{\rm CMB}$
takes the value $\delta _{b}/\delta _{c} \sim 3.0\times 10^{-3}$ as computed by running the Boltzmann code CAMB \citep{camb} for a $\Lambda {\rm CDM}$ cosmology and for 
the WMAP7 parameters adopted in the present study as listed in Table~\ref{tab:parameters}.
\begin{figure}
\includegraphics[scale=0.49]{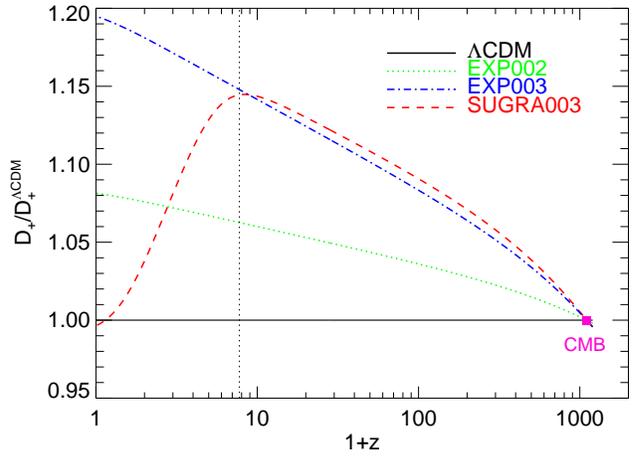}
\caption{The ratio of the linear growth factor $D_{+}$ of the different cosmologies considered in this work to the standard $\Lambda $CDM case. As the figure shows,
all the models share the same perturbations amplitude at $z_{\rm CMB}$, while their subsequent growth significantly differs from model to model. The most noticeable feature
of this evolution,
which constitutes one of the central outcomes of this work, is that the SUGRA cDE model, differently from the standard exponential cDE scenarios, features a clear
peak in the deviation of its growth factor from $\Lambda $CDM at $z_{\rm inv}\sim 6.8$, while it recovers the same amplitude of linear fluctuations as $\Lambda $CDM
at $z=0$.}
\label{fig:growth_factors}
\end{figure}
\normalsize
The result of the integrations is shown in Fig.~\ref{fig:growth_factors}, where the ratio of the total growth factor $D_{+}$ of each model over the standard $\Lambda $CDM case
is plotted as a function of redshift. As one can see, all the models start from the same amplitude of density perturbations at $z_{\rm CMB}$, but show 
a significantly different growth afterwards.

The standard exponential cDE model EXP002 and EXP003 (dotted green and dot-dashed blue lines, respectively) feature a faster growth during
the whole expansion history of the universe, reaching at $z=0$ an amplitude of density perturbations exceeding the $\Lambda $CDM prediction by 
$\sim 8 \%$ and $\sim 20 \%$, respectively. 
This is a very well known effect that has been widely discussed in the literature \citep[see \eg][]{Amendola_2000,Pettorino_Baccigalupi_2008,Baldi_etal_2010}, and that arises as a combination of the modified background expansion of the universe and of
the extra friction and fifth-force terms in Eq.~(\ref{gf_c}) that contribute to accelerate the growth of density perturbations. Such enhanced structure formation
process has been shown to account for an early formation of massive clusters \citep{Baldi_Pettorino_2011} but the faster growth until the present time also necessarily implies a significantly
higher normalization for the linear matter power spectrum at $z=0$, with a value of $\sigma _{8}=0.875$ for EXP002 and $\sigma _{8} = 0.976$ for EXP003, 
as compared to the assumed $\sigma _{8} = 0.809$ 
for the $\Lambda $CDM cosmology. 
Such a high normalization at $z=0$ could be in contrast with present measurements of the matter power spectrum at large scales \citep[see \eg][]{Montesano_Sanchez_Phleps_2011}
and represents the main unaddressed problem when trying to explain the existence of high-z massive clusters in the context of standard cDE models.
\ \\

The central result of the present work is that such problem might be avoided by bouncing cDE models, and in particular by the SUGRA cDE
scenario discussed in this paper. In Fig.~\ref{fig:growth_factors} the dashed red line shows the evolution of the ratio $D_{+}/D_{+}^{\Lambda {\rm CDM}}$
for our SUGRA003 cosmology. While at early times the growth proceeds in a similar fashion as for the standard exponential cDE model EXP003, 
this trend is suddenly stopped and inverted when the scalar field $\phi $ changes its direction of motion at $z_{\rm inv}$.
After this redshift the growth of density fluctuations is slowed down below the rate of the $\Lambda $CDM cosmology such that the discrepancy between the two models is progressively reduced and the amplitude of perturbations reaches again the $\Lambda $CDM value at $z=0$. This implies that the linear power spectrum
normalization of the two models will be the same both at $z_{\rm CMB}$ and at $z\sim 0$, while a significantly larger amplitude is reached in the SUGRA model
at an intermediate redshift $z_{\rm inv}$.

This peculiar evolution of the matter growth factor could not be realized in the context of standard gravity models or by a modification of the statistical properties of the
 initial conditions for the density fluctuations as for the case of non-Gaussian cosmological scenarios. It is therefore a quite specific footprint of a non trivial dynamical behavior 
of the DE scalar field at relatively late cosmological epochs. 
As we will show in the next Section, this feature of the SUGRA cDE model can account for a higher number density of massive clusters at 
early times while not significantly affecting the halo abundance at $z=0$.

Furthermore, since the different growth of structures induced by the DE interaction is expected to affect in a non-trivial way also the peculiar velocity field \citep[see \eg][]{Lee_Baldi_2011}, a detailed investigation of the impact of cDE models on redshift-space distortions in the large-scale distribution of galaxies might provide an independent way to test the cDE scenario \citep[][]{Marulli_Baldi_Moscardini_2011}.

\section{Non-linear structure formation}
\label{sec:sims}

In order to test directly the effects that the SUGRA cDE scenario has on the expected number density of massive halos at different cosmic epochs
we need to rely on the results of large N-body simulations which should include in their numerical treatment a self-consistent implementation
of the cDE effects described in Section~\ref{sec:cDE}.
To this end we make use of the publicly available halo catalogs of the {\small CoDECS} simulations suite\footnote{The {\small CoDECS} simulations are publicly available at the URL: www.marcobaldi.it/research/CoDECS} 
\citep[][]{CoDECS} which includes the very same
EXP002, EXP003 and SUGRA003 models investigated in the present work.
The {\small CoDECS} simulations have been performed using the modified version by \citet{Baldi_etal_2010} of the parallel Tree-PM N-body code
{\small GADGET-2} \citep{gadget-2}, that implements all the characteristic features of cDE models described above.
We refer to the {\small CoDECS} website for an extensive description of the numerical setup of the simulations, and we recall here only their main
features.

The $\Lambda $CDM-L, EXP002-L, EXP003-L, and SUGRA003-L simulations of the {\small CoDECS} database that are considered in the present work consist
of a cosmological box with a size of $1$ comoving Gpc$/h$ and periodic boundary conditions, filled with $1024^{3}$ CDM particles with a mass of 
$m_{c}=5.84\times 10^{10}$ M$_{\odot}/h$, and with $1024^{3}$ baryonic particles with a mass of 
$m_{b}=1.17\times 10^{10}$ M$_{\odot}/h$. 

Initial conditions are generated by perturbing a homogeneous {\em glass} distribution \citep{White_1994} in order to set up a random-phase realization
of the initial linear matter power spectrum which is taken to be the spectrum computed by the publicly available code CAMB \citep[][www.camb.info]{camb}
at the starting redshift of the simulations $z_{i}=99$ for the WMAP7 parameters specified in Table~\ref{tab:parameters}.
A common normalization of the linear perturbations amplitude at $z_{\rm CMB}$ consistent with present WMAP7 constraints is imposed to all the simulations
by properly rescaling the particles displacement between $z_{\rm CMB}$ and $z_{i}$ with the specific growth factors numerically computed for each cosmology.
This procedure makes the {\small CoDECS} runs more suitable to realistically quantify
the impact of cDE on the halo mass function with respect to previous attempts \citep[as \eg by][]{Baldi_Pettorino_2011}  
where a common normalization at the starting redshift of the simulations was assumed. 
A more accurate quantitative determination of the number counts deviation from $\Lambda $CDM can therefore be expected by using the {\small CoDECS}
halo catalogs. These are obtained by running a Friends-of-Friends (FoF) algorithm with linking length $\ell = 0.2 \times \bar{d}$, where $\bar{d}$ is the mean interparticle
separation. Since the {\small CoDECS} runs include both CDM and baryonic particles, the halo catalogs are compiled by running the FoF algorithm
on the CDM particles as primary tracers and then linking the baryonic particles to the FoF group of their closest CDM neighbor. 
It is also important to stress here that all the simulations are started with the same initial linear transfer function and with the same random seed for the realization of the matter
power spectrum in the initial conditions.
\begin{figure}
\includegraphics[scale=0.45]{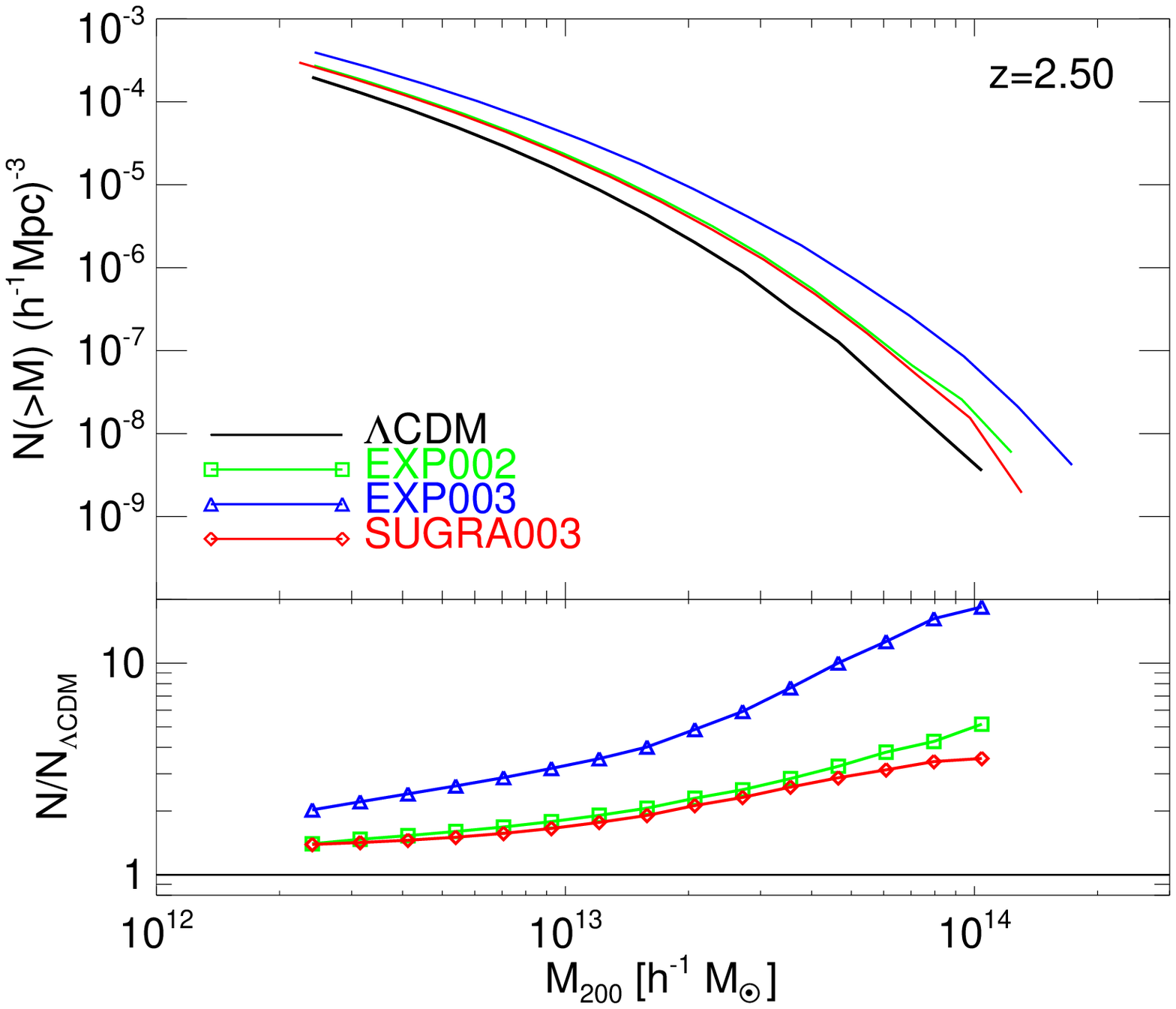}\\
\includegraphics[scale=0.45]{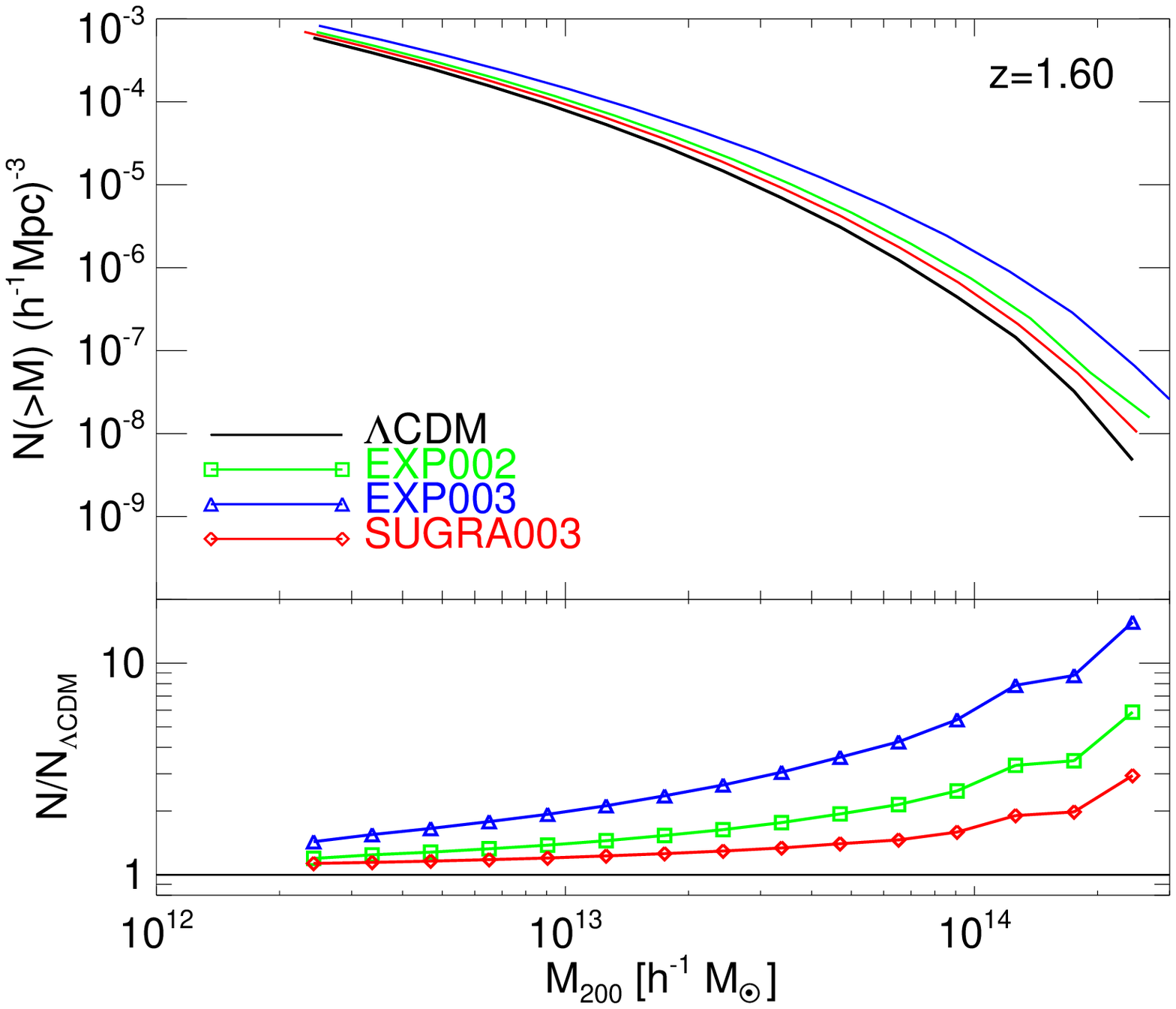}\\
\includegraphics[scale=0.45]{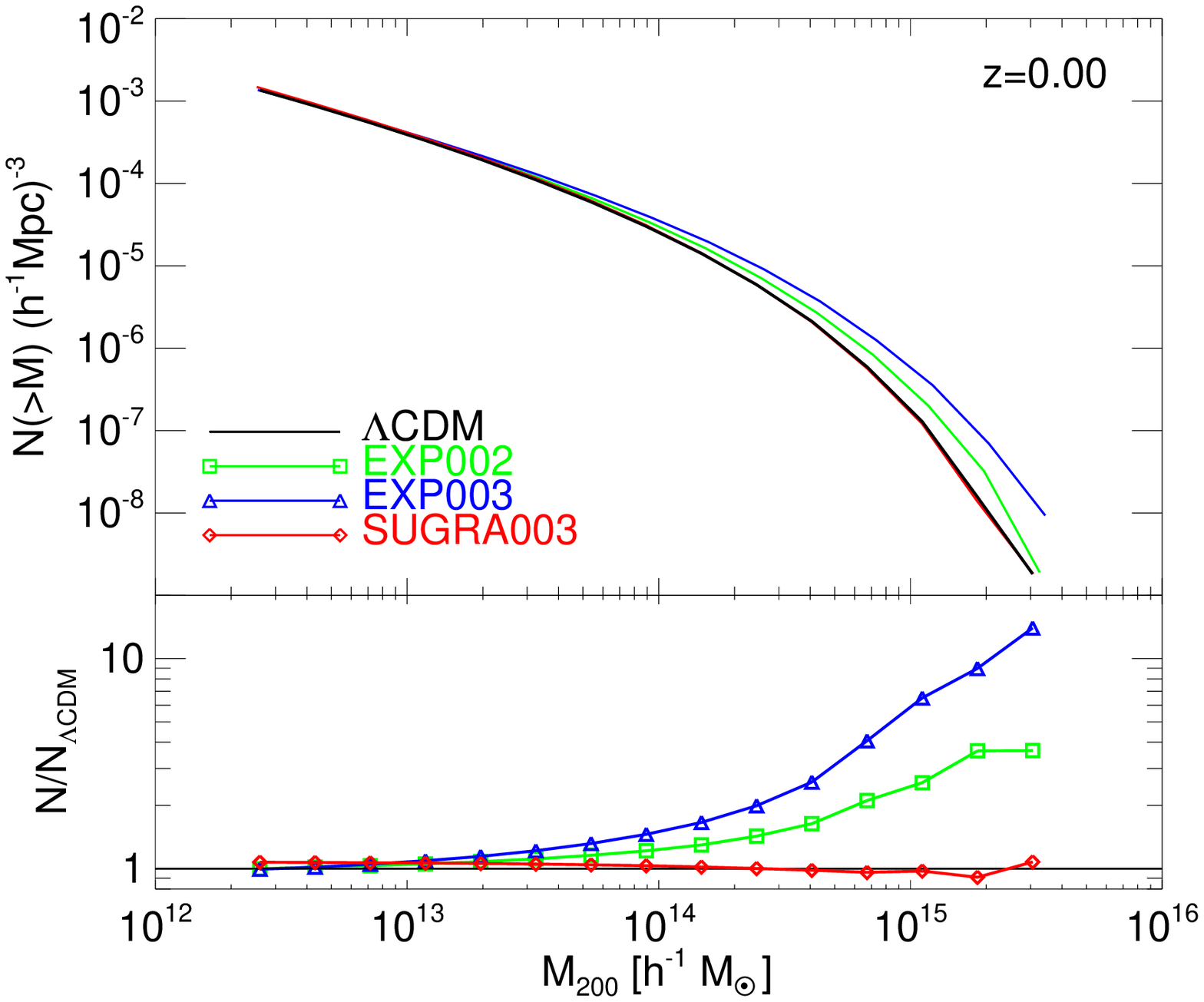}
\caption{The halo mass function as extracted from the halo catalogs of the {\small CoDECS} database, at $z=2.5$ (top), $z=1.6$ (middle), and
$z=0$ (bottom). For each panel, the upper plot shows the cumulative halo mass function in the mass range covered by the sample at that redshift, while
the lower plot shows the enhancement of the expected number counts with respect to $\Lambda $CDM. While all the cDE models predict a larger
number of massive halos at $z>0$, our proposed SUGRA cDE scenario is the only one that recovers the $\Lambda $CDM expectations at the present epoch, 
thereby avoiding tension with available bounds on the cluster abundance at $z=0$.}
\label{fig:massfunction}
\end{figure}
\normalsize

Having at hand all the {\small CoDECS} FoF halo catalogs, it is possible to compute the cumulative halo mass function for each of the models under discussion
at different redshifts and to compare their relative evolution. This is shown in the three panels of Fig.~\ref{fig:massfunction} for three choices of redshift, namely
$z=2.5$, $z=1.6$ and $z=0$. 
For convenience, the mass functions are plotted by binning halos into 15 logarithmically equispaced mass bins over the whole mass range covered by the sample
at each redshift.
\ \\

As one can clearly see from the Figure, at redshifts as high as $z=2.5$ all the cDE models show a larger number density of halos with respect to
$\Lambda $CDM over the whole mass range covered by the catalogs. This effect was already shown in the past by \citet {Baldi_Pettorino_2011}
but the results of the {\small CoDECS} simulations significantly extend its validity to higher masses, besides providing a more realistic quantitative
determination of the overall effect thanks to the common CMB normalization of the linear amplitude in all the models. The increase in the halo number density,
displayed in the bottom plot of each panel as the ratio of the halo mass function of each model over the $\Lambda $CDM case, shows a clear mass
dependence, being more significant at large masses and suggesting a possible further extrapolation of these results to more massive halos than the ones 
found in the {\small CoDECS} catalogs. In particular, the EXP003 model shows an increase of almost a factor of $20$ at $M\sim 1.0\times 10^{14}$ M$_{\odot}/h$
while at this redshift the EXP002 and SUGRA003 models determine a comparable increase of a factor $\sim 4-5$ at the same mass.
An accurate fit of the mass function, that we defer to an upcoming publication \citep{Cui_etal_2011}, will clearly be required in order to determine a reliable 
extrapolation of these numerical results to larger masses and to investigate in detail
to which extent the deviation in the number counts is degenerate with the evolution of the linear growth factor. 
Nevertheless, a rough estimate suggests that both the EXP002 and the SUGRA003 cosmologies
might reach at $z=2.5$ an enhancement factor of $\sim 10$ with respect to the $\Lambda $CDM halo mass function at $M\approx 3-6\times 10^{14}$ M$_{\odot}/h$.

At later times, as one can see from the central panel referring to $z=1.6$, the situation is already significantly different. 
While the EXP003 model still provides the strongest effect, reaching a factor of $\sim 10$ in the increase of the halo mass function
at $M\approx 2\times 10^{14}$ M$_{\odot}/h$, the  EXP002 and SUGRA003 models
now show a significantly different amplitude of the number counts enhancement: the former reaches an increase of a factor $\sim 6$ at 
$M\approx 2.5\times 10^{14}$ M$_{\odot}/h$, while a factor of $\sim 3$ is attained by the latter at the same halo mass. Also in this case, the enhancement
shows a clear mass trend in all the models, and a larger effect can be expected for halo masses beyond the range covered by the {\small CoDECS}
catalogs. 
\ \\

The central result of the present work is shown by the direct comparison of the top and central panels of Fig.~\ref{fig:massfunction} just discussed 
with the bottom panel,
that displays the halo mass functions of the different models at $z=0$. While both the standard exponential cDE models EXP002 and EXP003 still
show at the present time a similar enhancement of the halo number density as found at higher redshifts, with an increase at
$M\approx 3.0\times 10^{15} $M$_{\odot}/h$ of a factor of $\sim 4$ and $\sim15$, respectively, the bouncing cDE model SUGRA003 does not
show any excess in the halo number density over the whole mass range of the sample,  
and is therefore found to be perfectly consistent with the $\Lambda $CDM predictions on the cluster abundance at $z=0$.  This is a very remarkable result,
since it shows how the dynamic degree of freedom associated with a cDE scalar field $\phi $ can provide, for suitable self-interaction potentials, a natural
mechanism to explain an excess of massive clusters at high redshift as compared to the $\Lambda $CDM expectations without affecting the very tightly
constrained cluster mass function at the present time. Such behavior could not easily arise in the context of standard gravity theories or non-Gaussian
cosmological scenarios, and would therefore represent a clear indication of a dynamical nature of DE. 

Further detections of anomalously massive clusters at high redshift could therefore be considered as a ``smoking gun" for the 
existence of a dynamic degree of freedom in the dark sector.

\section{Conclusions}
\label{sec:concl}

In the present paper we have proposed a new model of interacting dark energy based on a SUGRA self-interaction potential
and on a constant coupling between dark energy and CDM particles. The most relevant feature of the SUGRA
self-interaction potential for the analysis carried out in this work, is the presence of a global minimum that allows the dark energy scalar field
to oscillate, thereby changing its direction of motion during the expansion history of the universe.
Such feature makes our SUGRA coupled dark energy scenario significantly different from previously proposed models of dark energy interactions,
as it allows an inversion of the energy flow between the two interacting components at relatively recent cosmological epochs. 

We have therefore compared the cosmological evolution of standard coupled dark energy models based on an exponential self-interaction potential
to our proposed SUGRA cDE scenario, both concerning the background dynamics and the evolution of linear and nonlinear perturbations. For the latter
task, we have also made use of large N-body simulations results through the publicly available {\small CoDECS} halo catalogs that include the specific models of dark energy interaction under investigation. 
\ \\

Already at the background level, the SUGRA cDE scenario shows a series of interesting features. First of all, the inversion of the direction of motion of the dark energy scalar field
happens in our specific model at $z_{\rm inv}\sim 6.8$ which is a redshift already relevant for astrophysical processes and for the early stages of nonlinear structure  formation. 
This inversion of motion 
corresponds to a ``bounce" of the dark energy equation of state parameter $w_{\phi }$ on the cosmological constant barrier $w_{\phi }=-1$, and imprints a specific 
pattern in the low-redshift evolution of the Hubble function that might provide a way to directly constrain the model. Furthermore, the inversion of the energy flow between
dark energy and CDM particle during cosmic evolution implies, for the specific model presented here, that the CDM particles mass has very similar values at the redshift
of the CMB and at the present time, thereby avoiding any mismatch of the cosmological parameters between $z_{\rm CMB}$ and $z=0$.

The most significant result of the present work, however, concerns the evolution of perturbations and the formation of linear and nonlinear structures in the context
of the SUGRA cDE scenario, as this is shown to provide appealing features to account for possible anomalous detections of massive clusters at high redshifts.
\ \\

At the linear level, we have demonstrated for the first time by numerically solving the linear perturbations equations that a bouncing 
coupled dark energy model, and in particular the SUGRA cDE scenario presented here, can determine the same amplitude of linear perturbations as a corresponding $\Lambda$CDM
cosmology both at CMB and at the present time, while featuring a significant deviation from $\Lambda $CDM at intermediate redshifts. In particular, both
the amplitude of scalar perturbations at CMB and the value of $\sigma _{8}$ at $z=0$ are the same between the standard $\Lambda $CDM model and our proposed SUGRA cDE
scenario, while the latter features a significantly larger amplitude of density perturbations at intermediate redshifts, with a peak in correspondence of the dark energy
``bounce" at $z_{\rm inv}$. Therefore, the SUGRA cDE model can be simultaneously consistent with CMB and local constraints on the perturbations amplitude, 
while allowing for significant freedom at intermediate epochs.
\ \\

We have then studied the evolution of the halo mass function as predicted by large N-body simulations that include all the characteristic features of the different interacting dark energy models considered in this work. Consistently with what found in previous studies, at high redshifts ($z\sim 2.5$) all the coupled dark energy models show a larger number of halos 
with respect to $\Lambda $CDM over the whole mass range covered by our numerical sample. The enhancement in the halo number density 
reaches a factor of $\sim 20$ at $M\sim 1.0\times 10^{14}$ M$_{\odot}/h$ 
for the most extreme standard coupled dark energy scenario, while the SUGRA cDE model shows an enhancement of a factor $\sim 4-5$ at the same mass. 
Furthermore, the enhancement has a clear mass dependence 
and increases towards larger masses, suggesting that the effect might be significantly larger at masses not covered by the halo sample of our simulations.
The subsequent evolution of the halo mass function clearly displays the fundamental difference between the standard coupled dark energy models based
on a monotonic self-interaction potential and the bouncing coupled dark energy scenarios, as the SUGRA cDE proposed here. At $z\sim 1.6$, in fact, all the models
still feature a significant excess of massive halos as compared to $\Lambda $CDM, although the SUGRA cDE model has reduced its enhancement factor much more
significantly than the other coupled dark energy models. Finally, at $z=0$ the SUGRA cDE model fully recovers the standard $\Lambda $CDM mass function over the whole
mass range of our sample, while the other coupled dark energy models still show a large excess of massive halos, 
especially in the range of very massive galaxy clusters, being therefore potentially in tension with available constraint on the cluster number counts.

This result, which represents the main outcome of the present paper, demonstrates for the first time by means of large and fully self-consistent N-body simulations
that coupled dark energy models with a suitable choice of the self-interaction potential can simultaneously account for the detection of anomalously massive
clusters at high redshifts and for the observed halo mass function at low redshifts. Such peculiar evolution could not arise in other types of models that have been
recently invoked as a possible explanation of the unexpected detection of massive clusters at high redshifts, as \eg non-Gaussian cosmological scenarios,
where an enhanced number density of halos at high redshifts necessarily implies a corresponding enhancement at low redshifts. 
\ \\

To conclude, we have studied a new class of interacting dark energy cosmologies characterized by a ``bounce" of the dark energy scalar field that is allowed to 
invert its direction of motion during the cosmic expansion. We have shown by means of linear and nonlinear numerical treatments that this new
class of models could be simultaneously consistent with observational constraints at CMB and at the present epoch, while allowing for significant deviations 
from the standard $\Lambda $CDM scenario at intermediate redshifts. In particular, we have shown this class of models to possess the (so far) unique feature of 
simultaneously accounting for unexpected detections of very massive clusters at high redshifts and
for the standard cluster abundance at the present time. Such behavior is a direct consequence of a non-trivial dynamics of dark energy at relatively
recent cosmological epochs, and therefore represents a specific observational signature of a possible dynamical origin of the accelerated expansion of the Universe.

\section*{Acknowledgments}

This work has been supported by 
the DFG Cluster of Excellence ``Origin and Structure of the Universe''
and by the TRR33 Transregio Collaborative Research
Network on the ``Dark Universe''.
I am deeply thankful to Valeria Pettorino for useful discussions on the models.
All the numerical simulations have been performed on the Power6 cluster at the RZG computing centre in Garching.

\bibliographystyle{mnras}
\bibliography{baldi_bibliography}

\begin{thebibliography}{63}
\expandafter\ifx\csname natexlab\endcsname\relax\def\natexlab#1{#1}\fi

\bibitem[Amendola(2000)]{Amendola_2000}
Amendola L., 2000, Phys. Rev., D62, 043511

\bibitem[Amendola(2004)]{Amendola_2004}
Amendola L., 2004, Phys. Rev., D69, 103524

\bibitem[Amendola et~al.(2008)Amendola, Baldi \&
  Wetterich]{Amendola_Baldi_Wetterich_2008}
Amendola L., Baldi M., Wetterich C., 2008, Phys. Rev., D78, 023015

\bibitem[Appleby \& Weller(2010)]{Appleby_Weller_2010}
Appleby S.~A., Weller J., 2010, JCAP, 1012, 006

\bibitem[Armendariz-Picon et~al.(2000)Armendariz-Picon, Mukhanov \&
  Steinhardt]{ArmendarizPicon_etal_2000}
Armendariz-Picon C., Mukhanov V.~F., Steinhardt P.~J., 2000, Phys. Rev. Lett.,
  85, 4438

\bibitem[Armendariz-Picon et~al.(2001)Armendariz-Picon, Mukhanov \&
  Steinhardt]{kessence}
Armendariz-Picon C., Mukhanov V.~F., Steinhardt P.~J., 2001, Phys. Rev., D63,
  103510

\bibitem[Astier et~al.(2006)]{SNLS}
Astier P., et~al., 2006, Astron. Astrophys., 447, 31

\bibitem[{Baldi}(2011{\natexlab{a}})]{Baldi_2011b}
{Baldi} M., 2011{\natexlab{a}}, Mon. Not. Roy. Astron. Soc., 414, 116

\bibitem[{Baldi}(2011{\natexlab{b}})]{CoDECS}
{Baldi} M., 2011{\natexlab{b}}, arXiv:1109.5695

\bibitem[{Baldi}(2011{\natexlab{c}})]{Baldi_2011a}
{Baldi} M., 2011{\natexlab{c}}, Mon. Not. Roy. Astron. Soc., 411, 1077

\bibitem[{Baldi} et~al.(2011{\natexlab{a}}){Baldi}, Lee \&
  Maccio]{Baldi_Lee_Maccio_2011}
{Baldi} M., Lee J., Maccio A.~V., 2011{\natexlab{a}}, Astrophys.J., 732, 112

\bibitem[{Baldi} \& {Pettorino}(2011)]{Baldi_Pettorino_2011}
{Baldi} M., {Pettorino} V., 2011, Mon. Not. Roy. Astron. Soc., 412, L1

\bibitem[{Baldi} et~al.(2010){Baldi}, {Pettorino}, {Robbers} \&
  {Springel}]{Baldi_etal_2010}
{Baldi} M., {Pettorino} V., {Robbers} G., {Springel} V., 2010, Mon. Not. Roy.
  Astron. Soc., 403, 1684

\bibitem[{Baldi} \& Viel(2010)]{Baldi_Viel_2010}
{Baldi} M., Viel M., 2010, Mon. Not. Roy. Astron. Soc., 409, 89

\bibitem[{Baldi} et~al.(2011{\natexlab{b}}){Baldi}, Pettorino, Amendola \&
  Wetterich]{Baldi_etal_2011a}
{Baldi} Marco M., Pettorino V., Amendola L., Wetterich C., 2011{\natexlab{b}},
  \mnras in press [arXiv:1106.2161]

\bibitem[Bean et~al.(2008)Bean, Flanagan, Laszlo \& Trodden]{Bean_etal_2008}
Bean R., Flanagan E.~E., Laszlo I., Trodden M., 2008, Phys. Rev., D78, 123514

\bibitem[Brax \& Martin(1999)]{Brax_Martin_1999}
Brax P., Martin J., 1999, Phys. Lett., B468, 40

\bibitem[Bremer et~al.(2006)]{Bremer_etal_2006}
Bremer M.~N., et~al., 2006, Mon. Not. Roy. Astron. Soc., 371, 1427

\bibitem[Brodwin et~al.(2010)]{Brodwin_etal_2010}
Brodwin M., et~al., 2010, Astrophys. J., 721, 90

\bibitem[Cayon et~al.(2010)Cayon, Gordon \& Silk]{Cayon_Gordon_Silk_2010}
Cayon L., Gordon C., Silk J., 2010, arXiv:1006.1950

\bibitem[{Cui} et~al.(in prep){Cui}, {Baldi} \& {Borgani}]{Cui_etal_2011}
{Cui} W., {Baldi} M., {Borgani} S., in prep

\bibitem[Ferreira \& Joyce(1998)]{Ferreira_Joyce_1998}
Ferreira P.~G., Joyce M., 1998, Phys. Rev., D58, 023503

\bibitem[Foley et~al.(2011)]{Foley_etal_2011}
Foley R.~J., et~al., 2011, Astrophys. J., 731, 86

\bibitem[Grossi et~al.(2007)Grossi, Dolag, Branchini, Matarrese \&
  Moscardini]{Grossi_etal_2007}
Grossi M., Dolag K., Branchini E., Matarrese S., Moscardini L., 2007, Mon. Not.
  Roy. Astron. Soc., 382, 1261

\bibitem[Holz \& Perlmutter(2010)]{Holz_Perlmutter_2010}
Holz D.~E., Perlmutter S., 2010, arXiv:1004.5349

\bibitem[Hoyle et~al.(2011)Hoyle, Jimenez \& Verde]{Hoyle_Jimenez_Verde_2011}
Hoyle B., Jimenez R., Verde L., 2011, Phys. Rev., D83, 103502

\bibitem[Hu \& Sawicki(2007)]{Hu_Sawicki_2007}
Hu W., Sawicki I., 2007, Phys. Rev., D76, 064004

\bibitem[Jee et~al.(2009)]{Jee_etal_2009}
Jee M.~J., et~al., 2009, Astrophys. J., 704, 672

\bibitem[Jee et~al.(2011)]{Jee_etal_2011}
Jee M.~J., et~al., 2011, arXiv:1105.3186

\bibitem[Jimenez \& Verde(2009)]{Jimenez_Verde_2009}
Jimenez R., Verde L., 2009, Phys. Rev., D80, 127302

\bibitem[Komatsu et~al.(2011)]{wmap7}
Komatsu E., et~al., 2011, Astrophys. J. Suppl., 192, 18

\bibitem[Kowalski et~al.(2008)]{Kowalski_etal_2008}
Kowalski M., et~al., 2008, Astrophys. J., 686, 749

\bibitem[La~Vacca et~al.(2009)La~Vacca, Kristiansen, Colombo, Mainini \&
  Bonometto]{LaVacca_etal_2009}
La~Vacca G., Kristiansen J.~R., Colombo L. P.~L., Mainini R., Bonometto S.~A.,
  2009, JCAP, 0904, 007

\bibitem[Lee \& {Baldi}(2011)]{Lee_Baldi_2011}
Lee J., {Baldi} M., 2011, arXiv:1110.0015, * Temporary entry *

\bibitem[Lewis et~al.(2000)Lewis, Challinor \& Lasenby]{camb}
Lewis A., Challinor A., Lasenby A., 2000, Astrophys. J., 538, 473

\bibitem[Li \& Barrow(2011)]{Li_Barrow_2011}
Li B., Barrow J.~D., 2011, Phys. Rev., D83, 024007

\bibitem[LoVerde et~al.(2008)LoVerde, Miller, Shandera \&
  Verde]{LoVerde_etal_2008}
LoVerde M., Miller A., Shandera S., Verde L., 2008, JCAP, 0804, 014

\bibitem[LoVerde \& Smith(2011)]{LoVerde_Smith_2010}
LoVerde M., Smith K.~M., 2011, arXiv:1102.1439

\bibitem[Lucchin \& Matarrese(1985)]{Lucchin_Matarrese_1984}
Lucchin F., Matarrese S., 1985, Phys. Rev., D32, 1316

\bibitem[Mantz et~al.(2010)Mantz, Allen, Rapetti \& Ebeling]{Mantz_etal_2010}
Mantz A., Allen S.~W., Rapetti D., Ebeling H., 2010, Mon. Not. Roy. Astron.
  Soc., 406, 1759

\bibitem[Marulli et~al.(2011)Marulli, Baldi \&
  Moscardini]{Marulli_Baldi_Moscardini_2011}
Marulli F., Baldi M., Moscardini L., 2011, arXiv:1110.3045

\bibitem[Matarrese et~al.(2000)Matarrese, Verde \&
  Jimenez]{Matarrese_Verde_Jimenez_2000}
Matarrese S., Verde L., Jimenez R., 2000, Astrophys. J., 541, 10

\bibitem[Montesano et~al.(2011)Montesano, Sanchez \&
  Phleps]{Montesano_Sanchez_Phleps_2011}
Montesano F., Sanchez A.~G., Phleps S., 2011, arXiv:1107.4097

\bibitem[Mortonson et~al.(2011)Mortonson, Hu \&
  Huterer]{Mortonson_Hu_Huterer_2011}
Mortonson M.~J., Hu W., Huterer D., 2011, Phys. Rev., D83, 023015

\bibitem[Mullis et~al.(2005)]{Mullis_etal_2005}
Mullis C.~R., et~al., 2005, Astrophys. J., 623, L85

\bibitem[Perlmutter et~al.(1999)]{Perlmutter_etal_1999}
Perlmutter S., et~al., 1999, Astrophys. J., 517, 565

\bibitem[Pettorino \& Baccigalupi(2008)]{Pettorino_Baccigalupi_2008}
Pettorino V., Baccigalupi C., 2008, Phys. Rev., D77, 103003

\bibitem[Ratra \& Peebles(1988)]{Ratra_Peebles_1988}
Ratra B., Peebles P. J.~E., 1988, Phys. Rev., D37, 3406

\bibitem[Reiprich \& Boehringer(2002)]{Reiprich_Boehringer_2002}
Reiprich T.~H., Boehringer H., 2002, Astrophys. J., 567, 716

\bibitem[Riess et~al.(1998)]{Riess_etal_1998}
Riess A.~G., et~al., 1998, Astron. J., 116, 1009

\bibitem[Rosati et~al.(2009)]{Rosati_etal_2009}
Rosati P., et~al., 2009, arXiv:0910.1716

\bibitem[Rozo et~al.(2010)]{Rozo_etal_2010}
Rozo E., et~al., 2010, Astrophys. J., 708, 645

\bibitem[Sartoris et~al.(2010)]{Sartoris_etal_2010}
Sartoris B., et~al., 2010, arXiv: 1003.0841

\bibitem[Sheth \& Diaferio(2011)]{Sheth_Diaferio_2011}
Sheth R.~K., Diaferio A., 2011, arXiv: 1105.3378

\bibitem[Simpson \& Bridle(2006)]{Simpson_Bridle_2006}
Simpson F., Bridle S., 2006, Phys.Rev., D73, 083001

\bibitem[Springel(2005)]{gadget-2}
Springel V., 2005, Mon. Not. Roy. Astron. Soc., 364, 1105

\bibitem[Tarrant et~al.(2011)Tarrant, van~de Bruck, Copeland \&
  Green]{Tarrant_etal_2011}
Tarrant E. R.~M., van~de Bruck C., Copeland E.~J., Green A.~M., 2011, arXiv:
  1103.0694

\bibitem[Vikhlinin et~al.(2009)]{Vikhlinin_etal_2009b}
Vikhlinin A., et~al., 2009, Astrophys. J., 692, 1060

\bibitem[Waizmann et~al.(2011)Waizmann, Ettori \&
  Moscardini]{Waizmann_Ettori_Moscardini_2011}
Waizmann J.~C., Ettori S., Moscardini L., 2011, arXiv: 1105.4099

\bibitem[Wetterich(1988)]{Wetterich_1988}
Wetterich C., 1988, Nucl. Phys., B302, 668

\bibitem[Wetterich(1995)]{Wetterich_1995}
Wetterich C., 1995, Astron. Astrophys., 301, 321

\bibitem[White(1994)]{White_1994}
White S. D.~M., 1994, arXiv: astro-ph/9410043

\bibitem[Xia(2009)]{Xia_2009}
Xia J.-Q., 2009, Phys. Rev., D80, 103514

\end{thebibliography}

\label{lastpage}

\end{document}